\documentclass[pre,aps,twocolumn,showpacs]{revtex4-1}
\usepackage{graphicx}
\usepackage{epsf}
\usepackage{amsmath}
\usepackage{amssymb}
\usepackage{xcolor}
\usepackage{mdframed}

\usepackage{enumerate}
\usepackage{nomencl}
\makeindex 
\makenomenclature
\renewcommand\nomgroup[1]{%
	\ifthenelse{\equal{#1}{A}}{%
		\item[\textbf{Acronyms}] }{
		\ifthenelse{\equal{#1}{R}}{%
			\item[\textbf{Roman Symbols}]}{
			\ifthenelse{\equal{#1}{G}}{%
				\item[\textbf{Greek Symbols }]}{
				\ifthenelse{\equal{#1}{S}}{%
					\item[\textbf{Superscripts  }]}{{
						\ifthenelse{\equal{#1}{U}}{%
							\item[\textbf{Subscripts }]}{{
								\ifthenelse{\equal{#1}{X}}{%
									\item[\textbf{Other Symbols }]}
								{{}}}}}}}}}}

\begin{document}
%
%
%
%
%
\title{Inferential framework for two-fluid model of cryogenic chilldown}


\author{DG Luchinsky$^*$, M Khasin}

\address{SGT Inc., Ames Research Center, Moffett Field, California}

\author{D Timucin}
\address{Ames Research Center, NASA, Moffett Field, CA, USA}

\author{J. Sass, B Brown}
\address{Kennedy Space Center, NASA, Kennedy Space Center, FL, USA}

		

\begin{abstract}
We report a development of probabilistic framework for parameter inference of cryogenic two-phase flow  based on fast two-fluid solver. We introduce a concise set of cryogenic correlations and discuss its parameterization. We present results of application of proposed approach to the analysis of cryogenic chilldoown in horizontal transfer line. We demonstrate simultaneous optimization of large number of model parameters obtained using global optimization algorithms. It is shown that the proposed approach allows accurate predictions of experimental data obtained both with saturated and sub-cooled liquid nitrogen flow. We discuss extension of predictive capabilities of the model to practical full scale systems.
\end{abstract}
		
\maketitle

		

	
	\section{Introduction}
	\label{s:intro}
	

Autonomous management of two-phase cryogenic flows is a subject of great interest to many spacefarers including effective human exploration of the Solar System~\cite{Chato08,Notardonato2012,Konishi:15}. It requires development of models that can recognize and predict cryogenic fluid dynamics on-line regime in nominal and faulty flow regimes without human interaction.

However, predicting the behavior of two-phase flows is a long standing problem of great complexity~\cite{Prosperetti07,Ishii10}. It becomes 
especially challenging when flowing fluids are far away from thermal equilibrium (e.g. during chilldown) and the analysis has to include heat and mass transfer correlations~\cite{TRACE50,Tong:97,Faghri:06,Ghiaasiaan:07}.

During past decades a number of efficient algorithms~\cite{TRACE50,RELAP5:1,Nourgaliev2012,relap-7} and advanced correlation relations for heat and mass transfer~\cite{TRACE50,SPACE:09,RELAP5:4} have been developed for analysis of multi-phase flows~\cite{Ishii10,Ghiaasiaan:07,Stadtke:06,WojtanII:05,Wojtan:06}. Despite this progress the state of the art in two-phase modeling lacks a general agreement regarding the fundamental physical models that describe the complex phenomena~\cite{relap-7}. As a consequence, uncertainties in modeling source terms may ultimately have a bigger impact on the results than the particular numerical method adopted~\cite{Prosperetti07}.

Analysis of cryogenic fluids introduces further complications due to relatively poor knowledge of heat and mass transfer correlations in boiling cryogenic flows~\cite{VanDresar:01,Umekawa:02,Jackson:06,Wang:13,Darr:15,Jason2015b}. Even less is known about flow boiling correlations of cryogenic fluids in microgravity~\cite{Konishi:15,Yuan:06}. To address these and other mission critical issues NASA has developed and implemented an impressive program of research, see e.g. ~\cite{Chato08,Notardonato2012,Kim2003,Robert2012}, that resulted in emergence of space based fluid management technologies.

Under this program a number of important experimental and modeling results have been obtained related to cryogenic two-phase flows (see e.g.~\cite{Konishi:15,Darr:15,Jason2015b,Kawaji1990,Kawaji1996,Majumdar2003,Majumdar2010,Chung:07,Chung2015} and references therein). Specifically, two-phase separated flow models were developed for some the flow regimes~\cite{Kawaji1990,Kawaji:93,Chung:09}. A number of optimization techniques have become commercially available for analysis of the model parameters and data correlations~\cite{Cullimore:98}.

However, small time steps and instabilities~\cite{Kawaji1990,Chung:09} or implicitness of numerical scheme~\cite{Majumdar2011b,SINDAFLUINT:5.6} impose substantial limitations on the speed of the solution, efficiency of multi-parametric optimization, and possibility of on-line application. As a result accurate predictions of transient cryogenic flows remain a challenge~\cite{Jason2015b,Cullimore:98} and extensive research is currently under way~\cite{Konishi:15,Darr:15}.

Some of the grand challenges of this analysis include inference of parameters of cryogenic correlations from experimental time-series data and extension of the results obtained from small experimental subsystems to full scale practical systems.

In this paper we report on the development of separated two-fluid model suitable for fast on-line analysis of cryogenic flows and introduce model-based inferential framework capable of efficient multi-parametric optimization of the model parameters. 

We demonstrate an application of this inferential framework to the problem of modeling chilldown in horizontal cryogenic line. This problem has been shown to be a difficult one to solve in the earlier research~\cite{Cullimore:98}. Using proposed approach we obtain accurate predictions for transient liquid nitrogen flow both under sub-cooled and saturated conditions.

The paper is organized as follows. In the next Section we briefly describe the model and algorithm of its integration. In the Section~\ref{s:prob_framework} we introduce probabilistic framework for inference of the model parameters, discuss the uncertainties in the source terms and their parameterization. In the Section~\ref{s:correlations} we introduced constitutive relations used to model source teams. The approach to the inference of model parameters is discussed in Section~\ref{s:inferring}. In the Section~\ref{s:transfer} we describe an application of the proposed technique to an analysis of cryogenic chilldown in horizontal pipe. Finally, in the Conclusions we summarize the obtained results and discuss directions of future work.

\section{Model}
\label{s:model}
	
We limit our analysis to one-dimensional flow networks having in mind fast on-line applications of the solver. To this end we have developed and tested a number of algorithms~\cite{LuchI:14a,LuchI:14b,LuchI:14d,Foygel:14,Foygel:15,Ponizovskaya:15,Luchinsky:15b,Luchinsky2015c}) to see if their speed and accuracy can satisfy requirements of real-time application. It was shown that the nearly-implicit algorithm, similar to one developed in~\cite{RELAP5:1}, can be applied successfully for on-line predictions of non-homogeneous ($u_g \neq u_l$) and non-equilibrium ($T_g\neq T_l$) flows.

In this section we will describe briefly the corresponding model equations and the algorithm of their integration. Extensive details can be found in~\cite{LuchI:14a,LuchI:14b,LuchI:16c,LuchI:14d}, see also~\cite{Luchinsky:15b,Luchinsky2015c}.

\subsection{Model Equations}
\label{ss:Equations}

In nearly implicit algorithm a closed system of equations is obtained assuming equal local pressure values for the both phases~\cite{Nourgaliev2012,Wallis:69,Stadtke:06}. The corresponding six-equation model consists of a set of conservation laws for the mass, momentum, and energy of the gas (see e.g.~\cite{RELAP5:1,Morales:12,TRACE50,LuchI:14a,LuchI:14b} )

\noindent
\begin{equation}\label{eq:CL-gas}
\hspace{-0.3cm}
\begin{array}{l}
{\left( {A{\alpha}{\rho _g}} \right)_{,t}} + {\left( {A{\alpha}{\rho _g}{u_g}} \right)_{,x}} = A{{\rm{\Gamma }}_g}\\
{\left( {A{\alpha}{\rho _g}{u_g}} \right)_{,t}} + {\left( {A{\alpha}{\rho _g}u_g^2} \right)_{,x}} + A{\alpha}{p_{,x}} = - A{\alpha}{\rho _g}{z_{,x}}\\
~~~~~~~~ - {\tau _{gw}}{l_{wg}} - {\tau _{gi}}{l_i} + A{{\rm{\Gamma }}_g}{u_{ig}}\\
{\left( {A{\alpha}{\rho _g}{E_g}} \right)_{,t}} + {\left( {A{\alpha}{\rho _g}{E_g}{u_g}} \right)_{,x}} = -Ap {\alpha_{,t}} - {\left( {pA\alpha {u_g}} \right)_{,x}} \\
~~~~~~~~ + {{\dot q}_{gw}}{l_{wg}} + {{\dot q}_{gi}}{l_i} + A{\rm\Gamma }_{g}H_{g}
\end{array}
\end{equation}
and liquid phases
\noindent
\begin{equation}\label{eq:CL-liquid}
\hspace{-0.3cm}\begin{array}{l}
{\left( {A\beta {\rho _l}} \right)_{,t}} + {\left( {A\beta {\rho _l}{u_l}} \right)_{,x}} =  - A{{\rm{\Gamma }}_g}\\
{\left( {A\beta {\rho _l}{u_l}} \right)_{,t}} + {\left( {A\beta {\rho _l}u_l^2} \right)_{,x}} + A\beta {p_{,x}} =  - A\beta {\rho _l}{z_{,x}} -\\
~~~~~~~~ {\tau _{lw}}{l_{wl}} - {\tau _{li}}{l_i} - A{{\rm{\Gamma }}_g}{u_{il}}\\
{\left( {A\beta {E_l}{\rho _l}} \right)_{,t}} + {\left( {A\beta {E_l}{\rho _l}{u_l}} \right)_{,x}} = -Ap {\beta_{,t}} - {\left( {pA\beta {u_l}} \right)_{,x}} +\\
~~~~~~~~ {{\dot q}_{lw}}{l_{wl}} + {{\dot q}_{li}}{l_i} - A{\rm\Gamma }_{g}H_{l}.\\
\end{array}
\end{equation}
Here $p$, $\alpha$, $T$, and $ \rho $ are pressure, temperature, and density of the fluid. $ E $ is the total specific energy, $ H_{g(l)}$ is the specific enthalpy of the gas generated (liquid evaporated)  at the interface and near the wall. $u$ is the fluid velocity, $\tau$ is the wall shear stress, and $\dot q$ is the heat flux at the wall and at the interface. The total mass flux $\Gamma_{g}=\Gamma_{wg}+\Gamma_{ig}$ has two components corresponding to the mass transfer at the wall $\Gamma_{wg}$ and at the interface $\Gamma_{ig}$.

The fluid dynamics equations are coupled to the equation for the wall temperature $T_w$

\noindent
\begin{equation}\label{eq:wall}
\begin{array}{l}{\rho _w}{c_w}{d_w}\frac{{\partial {T_w}}}{{\partial t}} = {h_{wg}}\left( {{T_g} - {T_w}} \right)\\ \qquad+ {h_{wl}}\left( {{T_l} - {T_w}} \right) + {h_{amb}}\left( {{T_{amb}} - {T_w}} \right).\end{array}
\end{equation}
Here $\rho$, $c$, and $d$ are density, specific heat, and thickness of the pipe wall, $h$ is the heat transfer coefficient corresponding to the ambient ($h_{amb}$) and internal heat flowing to the wall from the gas ($h_{wg}$) and liquid ($h_{wl}$) phases.

The characteristic feature of the model (\ref{eq:CL-gas}), (\ref{eq:CL-liquid}) is its non-hyperbolicity~\cite{Stadtke:06,Nourgaliev:03} related to the assumption of exclusively algebraic terms describing the interfacial drag and incomplete formulation for the interfacial momentum coupling.
It can be shown that this system does not have a complete set of real eigenvalues and does not represent a well-posed system of equations~\cite{LuchI:14a,Morales:12,Liou:09}.

It is also known that this system displays lack of positivity and instabilities due to phase appearance/disappearance process~\cite{Nourgaliev2012,Cordier:14}. In addition, the effect of algebraic source terms represents a system of ``stiff'' differential equations~\cite{Stadtke:06}
and roundoff errors may significantly contribute to numerical instabilities.


Despite these difficulties a number of algorithms were developed~\cite{TRACE50,RELAP5:1,Nourgaliev2012} and successfully employed to predict two-phase flows of boiling water in large scale system. In our development of the algorithm we were following the guidelines of earlier research.

\subsection{Algorithm}
	\label{ss:algorithm}
	
The choice of the algorithm was guided by the fact~\cite{Nourgaliev2012} that all current reactor thermal-hydraulics codes ~\cite{TRACE50,RELAP5:1,Bestion:90} originate from Liles and Reed~\cite{Liles:78} extension of Harlow and Asden~\cite{Harlow:68,Harlow:71} all-speed implicit continuous-fluid Eulerian algorithm. These codes enhance the stability of the method and eliminate material CFL restrictions using a couple of extensions: stability-enhancing two-step~\cite{TRACE50} and nearly implicit~\cite{RELAP5:1} algorithms.

In this work the discretization and integration of the model equations (\ref{eq:CL-gas}) - (\ref{eq:wall}) follow closely the nearly-implicit method described in RELAP5-3D~\cite{RELAP5:1} (see~\cite{LuchI:14b,Luchinsky:15b} for the details). The integration was performed in two steps.  The first  step of the algorithm can be briefly summarized as follows:
(i) Solve expanded equation with respect to pressure expressed in terms of new velocities;
(ii) Solve momenta equations written in the form of block tri-diagonal matrix  for the new velocities;
(iii) Find new pressure;
(iv) Find provisional values for energies and void fractions using expanded equations of states;
(v) Find provisional values of mass fluxes and heat transfer coefficients using provisional values of temperatures obtained.

At the second step new values of the densities, void fractions, and energies are found by solving the unexpanded conservation equations for the phasic masses and energies using provisional values for the heat and mass fluxes in source terms. The solution is reduced to independent solution of four tri-diagonal matrices. The values of pressure and velocities in these matrices are taken at the new time step.

The resulting scheme is efficient and fast and can integrate 1000 sec of real time chilldown in a few seconds of computational time. For a model consisting of $N$ control volumes it involves inversion of N $4\times 4$ matrices, solution of $2\times(N-1)$ tree-block-diagonal matrix equation, solution of four $N\times N$ tridiagonal matrix equations, and  $N\times m$ explicit computations. 

A special attention was paid to the stability of the code. Various methods are available for regularization of the solution including standard upwinding and staggered grid methods as well as ad hoc smoothing and multiple time step controls techniques, see~\cite{LuchI:14b} for the details. Specifically, multiple techniques can be used to suppress~\cite{RELAP5:1,Nourgaliev:03} the non-hyperbolicity.

In this work to suppress the non-hyperbolicty we are using so-called virtual mass term~\cite{RELAP5:1}
\[{M_V} = C\alpha \beta{\rho _m}\left[ {\frac{{\partial \left( {{u_g} - {u_l}} \right)}}{{\partial t}} + {u_l}\frac{{\partial {u_g}}}{{\partial x}} - {u_g}\frac{{\partial {u_l}}}{{\partial x}}} \right]\]
in the right hand sides of the momentum equations. In practical computations the terms corresponding to spatial derivatives were neglected.

The stability of the algorithm was further enhanced by using the time step control to insure that all the thermodynamic variables remain within the predetermined limits and that the change of these variables at any given time step does not exceed 25\% of their values obtained at the previous time step. If these conditions are not satisfied the time step is halved and integration is repeated. If time step goes beyond limiting value the integration is terminated. 

Similar control is applied to enforce mass conservation in each control volume and in the system as whole. In addition, smoothing mentioned above was found to be a very important tool to ensure stability of the scheme. In this work we followed recommendations provided by Liou~\cite{Liou:08} and adjust temperature, velocity, and density according to the following expression
\begin{equation}\label{eq:smoother}
{\phi _{adj}} = g(x){\phi _d} + \left( {1 - g(x)} \right){\phi _c},
\end{equation}
where
\[ g(x) = {x^2}\left( {2x - 3} \right);\quad {\rm and}\quad x = \frac{{{\alpha _d} - {x_{\min }}}}{{{x_{\max }} - {x_{\min }}}}.\]
Here ``d'' stands for disappearing phase and ``c'' for conducting phase. The exact values of the minimum and maximum void fraction $x_{\min}$ and  $x_{\max}$ were established using extensive numerical experimentation as$\sim 10^{-7}$ and $\sim 10^{-2}$ respectively. 


The set of equations (\ref{eq:CL-gas}), (\ref{eq:CL-liquid}), and (\ref{eq:wall}) is incomplete and a number of closure relations is required to close it. For cryogenic flows, however, the number of available experimental results is limited and further research is required to establish flow boiling correlations (see e.g.~\cite{VanDresar:01, Jackson:06}). 


It is, therefore, important that the cryogenic modeling is embedded within optimization framework that allows efficient inference of the correlation parameters and systematic comparison between various functional forms of the constitutive relations. 

We note that corresponding optimization framework is also one of the key tools required for autonomous control of cryogenic flows. Accordingly in the current work we were focused on development of an efficient optimization framework. Below we briefly outline this approach. 

\section{Probabilistic framework}
\label{s:prob_framework}

The most time consuming step in development of the cryogenic flow models is accurate correlation of the model predictions against experimental data (see e.g.~\cite{Ghiaasiaan:07} and references therein). This step is crucial for practical applications of the model including e.g. reliable design of cryogenic hardware~\cite{Cullimore:98}, analysis of nominal regimes of operation, fault detection and  isolation, and efficient recovery from off-nominal regimes. It becomes even more important when one attempts to extend model predictions to untestable conditions~\cite{Cullimore:98,Konishi2015} or to practical full-scale systems~\cite{Motil:04}.

At present the main approach to correlation of experimental data is based on fitting (mainly by hand) extensive databases obtained in various flow regimes under carefully controlled experimental conditions~\cite{TRACE50,RELAP5:6,Jason2015b}. The rationale behind this approach is an attempt to reduced a very large number of uncertainties inherent to the model and to obtain solution of the fitting problem using traditional techniques. 

However, such an approach becomes prohibitively expensive in development two-phase flow correlations and autonomous fluid management in microgravity. An efficient solution of the problem in this case has to rely on more intelligent and less expensive methods of inferring correlation parameters. This work is an attempt to establish feasibility of such methods. Below we briefly review the uncertainties of the model and outline probabilistic approach that can be applied to their analysis.

\subsection{Uncertainties }
\label{ss:uncertainties}


Fundamentally, the probabilistic nature of the model predictions is related to the fact that the interface between two phases is continuously fluctuating and neither location nor the shape of the interface can be resolved by the model. The spatial and time scales of these fluctuations are continuously changing depending on the flow regime. The intensity of these fluctuations is especially significant during chilldown, when liquid and vapor phases coexist under strongly non-equilibrium conditions, see e.g.~\cite{Chung:07}.

Another major source of uncertainty is related to the choice of the functional form of the correlations. There have been literally hundreds of various modifications  proposed for the flow boiling  correlations~\cite{Nellis:09,Shahs2006} and the corresponding functional space is continuously expanding~\cite{Darr:15,Kim2014}.

To illustrate this point let us consider as an example one of the key correlation parameters so-called critical heat flux,  $\dot q_{chf,0} $, corresponding to the maximum heat transfer from boiling fluid to the wall. 

One of the best known correlations for the pool boiling value of $\dot q_{chf,0} $ was proposed by 
Zuber~\cite{Zuber:58} in the form
\begin{equation}\label{eq:Zuber}
\dot q_{chf,0} = \frac{\pi}{24} h_{lg} \rho_g \left(\frac{\sigma g (\rho_l-\rho_g)}{\rho_g ^{2}}\right)^{1/4} \left(\frac{\rho_l }{\rho_l+\rho_g}\right)^{1/2}.
\end{equation}
\nomenclature[]{$h_{lg}$}{latent heat of evaporation}
Zuber's model assumes several approximations,  including e.g. the following: (i) rising jets with radius $R_j$ form a square grid with a pitch equal to the fastest growing wavelength due to Taylor instability, (ii) the rising jets are assumed to have critical velocity dictated by Helmholtz instability, (iii) the neutral wavelength of the rising jet is assumed to be $ 2\pi R_j $.

It is clear from the list of assumptions that numerical constants in Zuber's correlation have to  be viewed only as an approximation. Furthermore, this approximation does not take into account surface wettability, pipe curvature, sub-cooling, and surface orientation. Accordingly, several corrections are known~\cite{Ghiaasiaan:07} that modify functional form of this correlation. 

In boiling flows further corrections have to be introduced to take into account the dependence of the heat flux on the void fraction, velocity, and sub-cooling of the flow. For example, Griffith et all use the following functional form of the corresponding corrections for cryogenic flows~\cite{Franchello:93,Seader:65} 
\begin{eqnarray} 	
\label{eq:Griffith_chf}
\dot q_{chf}& =& \dot q_{chf,0} (\alpha_{cr} - \alpha)\left( 1 + a_1\left(\frac{\rho_l c_l \Delta T_{sub}}{\rho_g h_{lg}}\right)\right. \\
&& +\left.  a_2 Re_l + a_3\left(\frac{Re_l \rho_l c_l \Delta T_{sub}}{\rho_g h_{lg}}\right)^{1/2}\right), \nonumber
\end{eqnarray}
where  $\alpha_{cr}$ is the critical value of the void fraction and $a_i$ are constants, e.g. $a_1 = 0.0144$, $a_2 = 10^{-6}$, $a_3 = 0.5\times 10^{-3}$~\cite{Griffith:57}, and $\alpha_{cr} = 0.96$ ~\cite{Franchello:93} for water. Different functional forms of similar corrections are also known and will be considered below. 

In practice, we often used a simpler expression, cf~\cite{Franchello:93,Iloeje:82}
\begin{equation}\label{eq:Iloeje_chf}
\dot q_{chf} =\dot q_{chf,0}\cdot  a_1 \cdot (\alpha_{cr}-\alpha)^{a_2} (1+a_3 G^{a_5}),
\end{equation}
where typical values of parameters used in simulations are $a_1$=1.0, $\alpha_{cr}$=0.96, $a_2$=2.0, $a_3$=0.16, and $a_4$=0.2.

Another source of uncertainty is added to the problem by the fact that models are often correlated against multiple datasets obtained for different flow conditions. Some of these conditions (e.g. wettability) are not well known. 

As a result of multiple approximations the number of  parameters that have to be established in different flow regimes for practical full-scale systems is of the order of thousand. It becomes clear that computer base intelligent methods are required to handle complexity of this scale in realistic time frame.


\subsection{Probabilistic approach }
\label{ss:Probabilistic_approach}

Here we consider briefly one of the approaches to the solution of this problem based on  probabilistic Bayesian method~\cite{Ghahramani2015}. Using this technique one can~\cite{Ghahramani2015} estimate the probability of unknown model parameters 
\begin{equation}\label{eq:bayes_param}
	P(\theta|d,m) = \frac{P(d|\theta,m)P(\theta|m) }{P(d|m)},
\end{equation}
compare different models
\begin{equation}\label{eq:bayes_model}
	P(d|m) = \frac{P(m|d)P(m) }{P(d)}
\end{equation}
and forecast system response $d_{n}$ to untested experimental conditions
\begin{equation}\label{eq:bayes_forecast}
	P(d_{n}|d,m) =\int P(d_{n}|\theta,d,m)P(\theta|d,m) d\theta.
\end{equation}
\nomenclature[]{$M$}{model}
\nomenclature[]{$d$}{time-series data}
\nomenclature[G]{$\theta$}{set of model parameters}
\nomenclature[U]{$w$}{wall}

Here, $d$ is the experimental time-series data, $m$ is the model, and $\theta$ is the set of model parameters.

There are two important advantages of this approach to bear in mind. The first one is its ability to select simpler models over more complex models, thus resolving so-called ``Ockham's razor problem'' of optimization. The second advantage is its flexibility in the choice model parameters. It is known that the best predictive performance is often obtained~\cite{Ghahramani2015} using the most flexible system that can better adapt to the complexity of the data.

Accordingly, this approach allows for development of a flexible model with a set of parameters large enough to capture all the required properties of the data.

The main outcome of the method is distribution of the model parameters that contains maximum statistical information available in a given experimental data set. Importantly, this information can be updated when new time-series data or new database become available. As a result, the approach tends to provide the best fit to all available data. 


%
%
%
%
%

\subsection{Equivalent State Space Model}
\label{ss:SSM}

One of the key steps in developing probabilistic inferential framework is the solution of the optimization problem. In general terms this problem is formulated by presenting results of integration of equations (\ref{eq:CL-gas}) - (\ref{eq:wall}) 
on one time step $t$ in the form of discrete-time state-space model (SSM)
\begin{equation}\label{eq:SSM}
\begin{array}{ll}
x_{t+1} = f(x_t,c) + \varepsilon_t,\\
y_{t} = g(x_t,b) + \chi_t.
\end{array}
\end{equation} 
Here $c$ is the set of the model parameters and $x_t$ is a set of dynamical variables $ \left\lbrace \rho_g, \rho_l, T_g, T_l, u_g, u_l, p, \alpha \right\rbrace ^L_t $ at time $t$ discretized in space on a set of control volumes $V_L$.  

The observations $y_t$ in the SSM are related to the unobserved states $x_t$ via nonlinear function $g(x_t,b)$.
$\varepsilon_t$ and $\chi_t$ in equations (\ref{eq:SSM}) are independent identically distributed sources of Gaussian noise. The latter assumption is standard within this approach when sources of noise have multiple origin and are not well established (cf \cite{Ghahramani2015,Luchinsky2008,Duggento2009a}).

Although it is possible to determine simultaneously parameter and state of a systems within proposed framework (see e.g.~\cite{Ghahramani2015,Smelyanskiy2009,Duggento2009a}), here for simplicity we neglect measurement noise and assume that the key dynamical variables such as pressure $\hat{p}$, wall temperature $\hat{T}_w$, and fluid temperature  $\hat{T}_f$ can be measured directly in the experiment. This is indeed the case for the time-series data obtained during chilldown experiment at National Bureau of Standards~\cite{NBS:66} that will be considered below, see Sec.~\ref{s:transfer}.

In the simplest case of general importance the problem can be reduced to the curve fitting problem (cf ~\cite{Cullimore:98}). The model $m$ in this case is the set of equations (\ref{eq:CL-gas}) - (\ref{eq:wall}) completed with constitutive relations and equations of state for the liquid and gas~\cite{RELAP5:1,LuchI:14b}. Data $d$ correspond to the time-series data $\{\hat{p}, \hat{T}_f, \hat{T}_w\}$ of pressure, fluid and wall temperature obtained in the experiment. $\theta$ correspond to the model parameters that will be discussed in more details in the following section. 

The goal of the probabilistic approach is to use time-series data $d$ to update initial guess for the distribution of the model parameters $\theta$. Below we provide an example of development and application of this approach to the analysis of chilldown in cryogenic horizontal line.

\section{Constitutive relations}
	\label{s:correlations}

As was discussed above, the model (\ref{eq:CL-gas}) - (\ref{eq:wall}) has to be completed with  the equations of state and the constitutive relations. The equations of state can be included into the model in the form of NIST tables~\cite{LuchI:14b,NIST}. Functional and parametric form of constitutive relations, on the other hand, represent one of the main source of uncertainties in the model. The corresponding constitutive relations~\cite{LuchI:16c} define boundaries between flow and boiling regimes, interphase friction, the coefficient of virtual mass, wall friction, wall heat transfer, interphase heat and mass transfer. 

In practical calculations the boundaries between flow and boiling regimes have to be determined first. Frictional losses and coefficients for the heat and mass transfer are defined at the next step for each flow regime.

In this work the boundaries between flow regimes are estimated using simplified Wojtan et al~\cite{WojtanI:05} map. The map was simplified in two ways. First we used only a few transition boundaries as shown below. Next, we estimated the location of these boundaries in the coordinates of mass flow rate ($\dot m$) and vapor quality ($\chi$) using original expressions. Finally, we approximated the location of these boundaries using low-dimensional polynomials and used polynomial coefficients as fitting parameters.

The rationale behind approximate description of the boundaries of flow regimes is twofold. It is known~\cite{Jackson2006a} that Wojtan maps can only be considered as an approximation to the flow regimes for cryogenic fluids. In the experiments performed at NBS~\cite{NBS:66} the flow regimes were not established experimentally and therefore cannot be validated.

{\it Stratified-Wavy-to-stratified transition.} For the stratified to wavy stratified transition flow on the plain $(\dot m, \chi)$ we have
\[{\dot m_{strat}} = {\left\{ {\frac{{{{\left( {226.3} \right)}^2}{A_{ld}}A_{gd}^2{\rho_g}\left( {{\rho_l} - {\rho_g}} \right){\mu_l} g}}{{{\chi ^2}\left( {1 - \chi } \right){\pi ^3}}}} \right\}^{1/3}} + 20\chi,\]
\nomenclature[]{$A$}{cross-sectional area}
\nomenclature[]{$D$}{internal pipe diameter}
\nomenclature[]{$We$}{Weber number}
\nomenclature[]{$Fr$}{Froude number}
\nomenclature[]{$\dot m$}{mass flow rate}
\nomenclature[G]{$\chi$}{vapor quality}
\nomenclature[G]{$\mu$}{dynamic viscocity}
\nomenclature[U]{$ld$}{dimensionless quantity for liquid cross-section}
\nomenclature[U]{$gd$}{dimensionless quantity for gas cross-section}
where $A_{gd}$ and $A_{ld} $ are dimensionless cross-sectional area of the gas and liquid fractions.

{\it Stratified-Wavy-to-annular-intermittent transition.} The transition boundary from wavy-stratified to annular or intermittent flow is given  by the following relation
\[\begin{array}{*{20}{l}}
\dot m_{wavy} = \left\{ \frac{16 A_{gd}^3 g D \rho_l \rho_g}{\chi ^2 \pi ^2}\left[ 1 - \left( 2h_{ld} - 1\right)^2 \right]^{.5}\right.\\
\qquad\left.\times\left[ {\frac{{{\pi ^2}}}{{25 h_{ld}^2}}{{\left( {1 - \chi } \right)}^{ - {F_1}\left( q \right)}}\left( {\frac{{We}}{{Fr}}} \right)_l^{ - {F_2}\left( q \right)} + 1} \right] \right\}^{.5} + 50.
\end{array}\]
Here $We_l$ and $Fr_l$ are Weber and Froude numbers ($We_l = \frac{{\dot m_l^2D}}{{{\rho_l}\sigma }}$ and $Fr_l = \frac{{\dot m_l^2}}{{\rho_l^2 g D}}$), while $ h_{ld} $ is dimensionless height of the liquid level.
\nomenclature[]{$h_{ld}$}{dimensionless height of the liquid level}

{\it Dryout transition}  that takes into account heat flux from the wall has the form
  \[\begin{array}{*{20}{l}}
  \dot m_{dry} = \left[ 4.25\left( ln\left( {\frac{{0.58}}{\chi}} \right) + 0.52 \right)\left( {\frac{{{\rho_g}\sigma }}{D}} \right)^{.17}\right.\\
  \qquad\times\left.{{\left( g D{\rho_g}\left( {{\rho_l} - {\rho_g}} \right) \right)}^{.37}}{{\left( {\frac{{{\rho_l}}}{{{\rho_g}}}} \right)}^{.25}}{\left( \frac{1}{\tilde q} \right)}^{.7} \right]^{.93}.
  \end{array}\]

Here $\tilde q$ is the heat flux at the wall normalized by the characteristic heat flux corresponding to the departure from nucleate boiling in the form~\cite{Kutateladze:50}
\[ {q_{DNB}} = K\cdot\rho_g^{1/2}{h_{lg}}{\left[ {\sigma g\left( {{\rho_l} - {\rho_g}} \right)} \right]^{1/4}}.\]

%
%

The location of these boundaries determines transitions between various regimes of heat transfer and pressure losses as will be discussed briefly below.

	\subsection{Heat and mass transfer}
	\label{ss:heattransfer}

The total mass transfer $\Gamma_g$ in equations (\ref{eq:CL-gas}), (\ref{eq:CL-liquid}) is the sum of the mass transfer at the wall and at the interface
\[{\Gamma _g} = {\Gamma _{wg}} + {\Gamma _{ig}},\]
where
\[{\Gamma _{wg}} = \frac{{{\dot q}_{wl}}}{{H_g^* - H_l^*}};\quad  \quad {\Gamma _{ig}} = \frac{{{\dot q}_{li} + {{\dot q}_{gi}}}}{{H_g^* - H_l^*}};\\
\]
and
\[ H_g^* - H_l^* = \left\{ {\begin{array}{*{20}{c}}
	{{H_{g,s}} - {H_l},\quad \Gamma  > 0}\\
	{{H_g} - {H_{l,s}},\quad \Gamma  < 0}
	\end{array}} \right..\]
%
The heat transfer correlations are subject of extensive research~\cite{Tong:97,Faghri:06}.
Here we briefly outline a subset of these correlations selected in the present work. The heat fluxes at the wall and at the interface are defined as follows
\[\begin{array}{*{20}{l}}
{\dot q_{wg} = h_{wg}\left( {T_w - 
			 T_g} \right);\quad \dot q_{ig} = h_{ig}\left( {
			 T_{l,s} - 
			 T_g} \right);}\\
{\dot q_{wl} = h_{wl}\left( {T_w - 
		T_l} \right);\quad \dot q_{il} = h_{il}\left( {
		T_{l,s} - 
		T_l} \right).}
\end{array}\]

In the current work we are interested in relatively low mass fluxes, $G < 600 kg/m^2/s$. In this limit correlations for the heat flux are often based on the multiplicative or additive corrections to the values obtained for pool boiling~\cite{Franchello:93, Seader:65, Griffith:57, Zuber:58, Kutateladze:59}. The following heat transfer mechanisms are included in the analysis: (i) convection, (ii) nucleate boiling, (iii) transition boiling, (iv) film boiling, and (v) transition to dryout regime.


\subsubsection{Convective Heat Transfer}\label{ss:convective}

Convective heat transfer in horizontal pipes distinguishes four flow regimes: (F-L) forced laminar, (F-T) forced turbulent, (N-L) natural laminar, and (N-T) natural turbulent convection. The corresponding correlations for the convective heat transfer can be taken in the form e.g.~\cite{TRACE50,RELAP5:4}
\begin{equation}\label{eq:convection_Nu}
h_{cb} = \frac{\kappa}{D_h}\left\{ {\begin{array}{*{20}{l}}
{4.36,\qquad \qquad \qquad    ~ \hbox{\rm{F-L~~\cite{Nellis:09}}};}\\
{0.023\cdot R{e^{0.8}}P{r^{0.4}},\quad   ~ \hbox{\rm{F-T~~\cite{Nellis:09}}};}\\
{0.1\cdot (Gr\cdot Pr)^{1/3}},\quad   ~~ \hbox{\rm{N-L~~\cite{Holman:89}}};\\
{0.59\cdot (Gr\cdot Pr)^{1/4}},\quad   ~ \hbox{\rm{N-T~~\cite{Holman:89}}}.\end{array}} \right.
\end{equation}
Here $Pr=\frac{\mu C_p}{\kappa}$ and $Gr=\frac{\rho^2 g \beta_T (T_w - T_{l(g)})D^3}{\mu^2}$ are Prandtl and Grashof numbers respectively, $\beta_T$ is the coefficient of thermal expansion, and $D_h$ is the hydraulic diameter. To guarantee a smooth transition between various regimes the maximum value of $h_{cb}$ is taken as the value for the convective heat transfer.
\nomenclature[U]{$cb$}{convective boiling}
\nomenclature[U]{$s$}{satturation}
\nomenclature[U]{$sub$}{subcooled}
\nomenclature[]{$D_h$}{hydraulic pipe diameter}
\nomenclature[G]{$\kappa$}{thermal conduciviy}

We note that convective heat transfer in the stratified flow does not significantly affect the chilldown process, because the fluid temperature in this regime is close to (or lower than) saturation temperature $T_s$. The first critical temperature that defines  the shape of the boiling curve and influences the chilldown corresponds to the onset of nucleation boiling $T_{onb}$.

\subsubsection{Onset of nucleate boiling}\label{sss:par_onb}

The correlations for onset of nucleate boiling are based on the analysis of the balance between mechanical and thermodynamical equilibrium~\cite{Ghiaasiaan:07}. Using this analysis the  $T_{onb}$ and the corresponding heat flux $\dot q_{onb}$ can be written in the form~\cite{Sato:64,Frost:67,Huang:09}

\begin{eqnarray}
T_{onb} &=& T_s+F\left( {1 + \sqrt {1 + \frac{2 \Delta T_{sub}}{F}} } \right),\label{eq:T_onb}\\
\dot q_{onb} &=& \frac{B}{Pr^2}~\Delta T^2_{sat} = h_{cb}(T_{onb} - T_{l})\label{eq:q_onb}
\end{eqnarray}
where
$ B = \frac{{{\rho _g}{h_{lg}}{\kappa _l}}}{{8\sigma {T_s}}}$, $F = \frac{h_{cb} Pr_l^2}{2B}$,   $\Delta T_{sat}=T_{onb}-T_s$ is the wall superheat, and $\Delta T_{sub}=T_s-T_l$ is liquid subcooling temperature. The convective heat transfer coefficient is given by  (\ref{eq:convection_Nu}).
\nomenclature[U]{$onb$}{onset of nucleate boiling}%

When the wall superheat exceeds $\Delta T_{sat} = T_{onb} - T_s$ the nucleation boiling begins and the heat flux to the wall may increase by more than an order of magnitude significantly affecting the chilldown process. This increase continues until the heat flux approaches its critical value $\dot q_{chf}$.

\subsubsection{Critical heat flux}\label{sss:par_cht}

The values of critical heat flux $\dot q_{chf}$ and the corresponding critical wall superheat $T_{chf}$ are crucial for predicting chilldown and dryout phenomena in non-equilibriums flows. In nuclear reactor codes (see e.g.~\cite{RELAP5:4,TRACE50}) these values are determined using look-up tables based on extensive experimental measurements obtained under various flow conditions. For cryogenic fluids experimental data remain sparse and values of $\dot q_{chf}$ and $T_{chf}$ are often estimated using mechanistic models, see e.g.~\cite{Tong:97,Ghiaasiaan:07,Seader:65,Crowley:89}, see also~\cite{Konishi2015} for a recent review.

The correlations for the critical heat flux where introduced in Sec.~\ref{ss:uncertainties}. The temperature $T_{chf}$ for the critical heat flux was estimated in this work using approach proposed by Theler~\cite{Theler:11}
\begin{equation}\label{eq:theler}
T_{chf} = \frac{T_s}{1-\frac{T_s R_g}{h_{lg}}log(2 k_g +1)},
\end{equation}
where $k_g$ is the isoentropic expansion factor that for ideal diatomic gases is 7/2 and $R_g$ is the specific gas constant.

%
%
%
%

When wall superheat exceeds $\Delta T_{chf} = T_{chf} - T_s$, the transition boiling begins and the heat flux to the wall decreases sharply  as a function of the wall temperature until the latter reaches minimum film boiling temperature $T_{mfb}$.

\subsubsection{Minimum film boiling}\label{sss:par_mfb}

In the film foiling regime the fluid flow is completely separated from the wall by the vapor film. The minimum value of the wall superheat $\Delta T_{mfb} = T_{mfb} - T_s$ corresponding to this regime was estimated by Berenson as~\cite{Carbajo:85, Berenson:61}
\begin{equation}\label{eq:Tmfb}
\begin{array}{l} 	\Delta {T_{mfb,0}} = 0.127\frac{\rho _g h_{lg}}{\kappa _g}\times\\
\qquad\left[ {\frac{{g\left( {{\rho _l} - {\rho _g}} \right)}}{{{\rho _l} + {\rho _g}}}} \right]^{2/3} {\left[ {\frac{\sigma }{{g\left( {{\rho _l} - {\rho _g}} \right)}}} \right]^{1/2}}{\left[ {\frac{{{\mu _g}}}{{\left( {{\rho _l} - {\rho _g}} \right)}}} \right]^{1/3}}
\end{array}
\end{equation}
\nomenclature[]{$c_p$}{specific heat for constant pressure}
\nomenclature[U]{$fb$}{film boiling}

Iloeje~\cite{Franchello:93, Iloeje:82} has corrected Berenson equation to take into account the dependence of the $\Delta T_{mfb}$ on the quality and mass flux of the boiling flows in the form
\begin{equation}\label{eq:Iloeje}
  \Delta {T_{mfb}} = c_1 \Delta {T_{mfb,0}} (1-c_2 X_e^{c_3}) (1+c_4 G^{c_5}),
\end{equation}
where  $X_e$ is the equilibrium quality, $G$ is liquid mass flux and $a_i$ are constants, e.g. $a_1 = 0.0144$, $a_2 = 10^{-6}$, $a_3 = 0.5\times 10^{-3}$~\cite{Griffith:57}, and $\alpha_{cr} = 0.96$ ~\cite{Franchello:93} for water. 
\nomenclature[]{$G$}{mass flux}
\nomenclature[U]{$mfb$}{minimum film boiling}

The heat flux in the film boiling flow can be chosen following e.g. recommendations of Groeneveld and Rousseau~\cite{Groeneveld:83}. In this work the heat flux to the wall in the film boiling regime was taken in the form of Bromley correlations
\begin{equation}\label{eq:Bromley}
h_{br} = C \cdot {\left[ {\frac{{g{\rho _g}\kappa _g^2\left( {{\rho _l} - {\rho _g}} \right){{\tilde h}_{lg}}{c_{pg}}}}{{D\left( {{T_w} - {T_{spt}}} \right)P{r_g}}}} \right]^{0.25}},
\end{equation}
corrected using Iloeje-type correlations~\cite{Franchello:93, Iloeje:82}
\begin{equation}\label{eq:Iloeje_hfb}
h_{fb} = c_1 h_{br} (1-c_2 X_e^{c_3}) (1+c_4 G^{c_5})
\end{equation}
Typical values of the parameters used in simulations are the following: (i) $c_1 = 2.0$; (ii)  $c_2 = 1.04$; (iii) $c_3 = 2.0$; (iv) $c_4 = 0.2$; (v) $c_5 = 0.1$.

The minimum film boiling heat flux can now be defined as
\begin{equation}\label{eq:q_mfb}
\dot q_{mfb} = h_{fb} \Delta T_{mfb}.
\end{equation}

\nomenclature[U]{$wg$}{wall  to gas}
\nomenclature[]{$X$}{mass quality}
\nomenclature[U]{$e$}{equilibrium}

To complete the discussion of the boiling heat transfer we notice that in the region of single phase gas flow the  heat transfer is given by equations (\ref{eq:convection_Nu}) with appropriately modified parameters. Transition to the single phase heat transfer is initiated when dryout transition is detected.


 
\subsection{Parametrization}
\label{ss:parametrization}

It follows from the discussion above that boundaries between various flow boiling regimes are characterized by a number of critical points including onset of nucleate boiling, critical heat flux, minimum film boiling, and onset of dry-out. The heat flux to the wall at these points may differ by an order of magnitude. 

To simplify the analysis of correlations in two-phase flow-boiling regimes the corresponding values of the heat flux can be anchored to the values at critical points as follows.

In the regime of nucleate flow boiling 
when the wall superheat increases from $\Delta T_{onb}$ to $\Delta T_{chf}$. the heat flux can be defined using simple interpolation
\begin{equation}\label{eq:q_nb_int}
  \dot q_{nb} = y^n \dot q_{onb} + (1-y^n) \dot q_{chf},
  \end{equation}
where $n$ is constant, $y$ is defined as $(T_w-T_{onb})/(T_{chf}-T_{onb})$, while $\dot q_{chf}$ and $\dot q_{onb}$ are given by the equations (\ref{eq:Griffith_chf}) and (\ref{eq:q_onb}) respectively.

Similar correlations were applied to interpolate transition boiling in the form~\cite{TRACE50}
\begin{equation}\label{eq:transition_boiling}
\dot q_{tb} = f_{tb}\cdot \dot q_{chf}+(1-f_{tb})\dot q_{mfb},
\end{equation}
where
$f_{tb} = \left(\frac{T_w-T_{mfb}}{T_{chf}-T_{mfb}}\right)^2$, where $T_{chf}$, $T_{mfb}$, and $\dot q_{mfb}$ are given by equations (\ref{eq:theler}), (\ref{eq:Iloeje}), and (\ref{eq:q_mfb}) respectively. We note that $\dot q_{chf}$ is the same as in eq. (\ref{eq:q_nb_int}) and $T_{chf}$ was corrected using Iloeje-type correlations~\cite{Iloeje:82} similar to the one applied in eq. (\ref{eq:Iloeje}).

Within this approach the flow boiling correlations are essentially controlled by parameterization of the set of characteristic points on the boiling curve  $ c_{cr} = \{(T_{onb}, \dot q_{onb})$, $(T_{chf}, \dot q_{chf})$, $(T_{mfb}, \dot q_{mfb})$, $(T_{dry}, \dot q_{dry}) \}$. By introducing corrections to temperature and heat transfer coefficient for critical heat flux and minimum film boiling in the form (\ref{eq:Iloeje}) and (\ref{eq:Iloeje_hfb}) we were able to obtain smooth transformation of the boiling surface between the pool and flow boiling regimes.

\begin{figure}[h!]
	\centering
	\hspace{-0.25cm}\includegraphics[width=8.25cm,height=6.5cm]{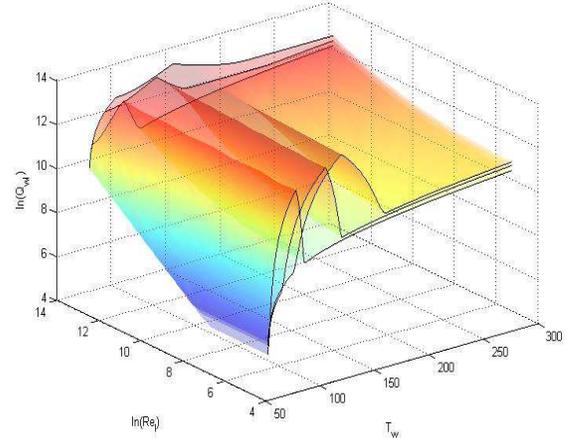}\\
	\caption{Heat flux from the liquid to the wetted wall as a function of the
		Reynolds number of the liquid flow and wall temperature $T_w$ calculated
		for three different pressures: 1, 3, and 7 atm. }\label{fig:Surface1}
\end{figure}

An example of such transformation is shown in Fig.~\ref{fig:Surface1}. In this figure the wall heat flux was calculated as a function of the wall temperature and Reynolds number of the liquid nitrogen for three different values of pressure.

	\subsection{Pressure drop}
	\label{ss:pressuredrop}
To complete the discussion of the constitutive relations, we briefly consider pressure drop correlations used in this research.

For the single phase flow the wall drag was calculated using following relations
\begin{equation}\label{eq:taitel76}
\tau_{wl} = f_{wl}\frac{\rho_l u_l^2}{2}, \quad\quad \tau_{wg} = f_{wg}\frac{\rho_g u_g^2}{2},
\end{equation}
Here the friction factors for turbulent and laminar flow are given by Churchill approximation
\noindent
\begin{equation}\label{eq:churchill}
f_{wg(l)}=2\left[\left(\frac{8}{Re}\right)^{12}+\frac{1}{\left(a+b\right)^{3/2}}\right]^{1/12},
\end{equation}
with Reynolds numbers
\[Re_{m,L} = \frac{\rho_{m,L}u_{m,L}D_{m,L}}{\mu_{g(l)}}\]
based on volume centered velocities $u_{m,L}$ and hydraulic diameter $D_m = \frac{4~A_L}{l_{m,L}}$
for each control volume. Index $m$ takes values $m=\{g,~l,~i\}$ for gas, liquid, and interface in a given control volume.

The coefficients $a$ and $b$ have the following form
\[
\begin{array}{l} 
a=\left\{ 2.475\cdot log\left[ \frac{1}{\left(\frac{7}{Re}\right)^{0.9}+0.27\left(\frac{\epsilon}{D_h}\right)}\right]\right\}^{16}, \\
b=\left( \frac{3.753\times 10^4}{Re}\right)^{16}.
\end{array}
\]

The two-phase friction pressure drop $\left( {\frac{{dp}}{{dz}}} \right)_{2\phi }$ is defined using Lockhart-Martinelli correlations \cite{Chisholm:67}. The pressure losses are partitioned between the phases as follows~\cite{RELAP5:1}
\[\begin{array}{l}
{\tau _{wg}}{l_{wg}} = {\alpha _g}{\left( {\frac{{dp}}{{dz}}} \right)_{2\phi }}\left( {\frac{1}{{{\alpha _g} + {\alpha _l}{Z^2}}}} \right),\\
{\tau _{wl}}{l_{wl}} = {\alpha _l}{\left( {\frac{{dp}}{{dz}}} \right)_{2\phi }}\left( {\frac{{{Z^2}}}{{{\alpha _g} + {\alpha _l}{Z^2}}}} \right).
\end{array}\]
Here $Z^2$ is given by
\[{Z^2} = {{\left( f_{wl}Re_l{\rho _l}u_l^2\frac{\alpha_{wl}}{\alpha_l} \right)} \mathord{\left/
		{\vphantom {{\left( {{f_l}R{e_l}{\rho _l}u_l^2\frac{{{\alpha _{wl}}}}{{{\alpha _l}}}} \right)} {\left( {{f_{wg}}R{e_g}{\rho _g}u_g^2\frac{\alpha _{wg}}{\alpha _g}} \right)}}} \right.
		\kern-\nulldelimiterspace} \left( f_{wg}R{e_g}{\rho _g}u_g^2\frac{\alpha _{wg}}{\alpha _g} \right)},\]
friction factor $f_{wg(l)}$ is in eq. (\ref{eq:churchill}), while coefficients $\alpha_{wl}$ and $\alpha _{wg}$ depend on the flow pattern~\cite{RELAP5:1}.

The interface drag is given by
\[{\tau _{ig}} =  - {\tau _{il}} = \frac{1}{2}{C_D}{\rho _g}\left| {{u_g} - {u_l}} \right|\left( {{u_g} - {u_l}} \right),\]
where interfacial drag coefficient $C_D$ depends on the flow pattern~\cite{TRACE50}.

We note that the functional form of the correlations adopted in this work is not unique and a number of alternative presentations can be used, see e.g.~\cite{RELAP5:4,TRACE50,Collier:94,LuchI:16c} for further details. 
The main goal of the present analysis is to develop an efficient approach to the parameter inference and systematic comparison between alternative functional forms of these correlation. 

\section{Inference of the model parameters} 
\label{s:inferring}

The discussion in previous sections has emphasized the fact that modeling of cryogenic flows involves a large number of unknown parameters. We will now show that proposed probabilistic framework allows for their efficient simultaneous estimation. 

The following steps are included into the process: (i) choice of the model parameters; (ii) definition of the objective (cost) function; (iii) estimation of the initial distribution of the model parameters via sensitivity study; (iv) simplified direct search for approximate globally optimized parameter values; (v) refined estimation of the optimal parameter values using global optimization; and (vi) estimation of the variance of the model parameters.

\subsection{Model parameters}
\label{ss:sys_parameters}

Analysis of the correlations of the two-phase boiling flows in full scale industrial systems may involve hundreds of model parameters~\cite{RELAP5:4}. In the present simplified model of the chilldown in horizontal straight line we limited studies to a set of 47 parameters divided into several groups, including e.g. parameters for: (i) onset of nucleate boiling; (ii) critical heat flux; (iii) film boiling; (iii) convective heat transfer; (iv) flow regime boundaries; and (v) frictional losses.


For example, parameters related to the Iloeje's corrections (\ref{eq:Iloeje_hfb}) to the minimum film boiling temperature  $ T_{mfb} $ are combined in a group shown in Table ~\ref{t:p_cor_Tmfb}.
Similar subsets of parameters were formed for other groups, see \cite{LuchI:16c} for further details.

\begin{table}[h!]
	\caption{\label{t:p_cor_Tmfb} Example of parameters for the temperature $ T_{mfb} $.  } 
	\vspace{-0.5cm}
	\begin{center} 
		\begin{tabular}{*{3}{l}}
			Parameter  &  Comment\cr 
			\hline
			Tmfbsc = 1.25; & \% overall scaling coefficient\cr
			Gtmfbsc = 0.2; &  \% scaling of the mass flow rate  \cr 
			Etmfbsc  = 0.165; &  \% exponent of the mass flow rate  \cr 
			Xetmfbsc = 2.0; &  \% scaling of the void fraction  \cr 
			Xtmfbsc  = 1.04; &  \% exponent of the void fraction  \cr 
			\hline
		\end{tabular}
	\end{center}
\end{table}

Not all the parameters are equally important/sensitive for the system dynamics. Relative significance of the model parameters depends strongly on the objective of optimization, the stage of the chilldown process, and the location of the sensors in the system. Accordingly, the first step in the analysis of the sensitivity is an appropriate choice of the objective function. 

	\subsection{Cost function}
	\label{ss:cost}
	
The primary goal of modeling large scale cryogenic systems is the ability to reproduce and predict system response in a variably of nominal and off-nominal regimes. The natural choice of the objective in this case is to minimize the sum of square difference between model predictions (${x}^k_{n}$) and data ($\hat{x}^k_n$) measured by different types of sensors at various locations. 

Typically, the fluid and wall temperatures and the fluid pressure are available for the measurements during chilldown. Taking into account time discretization of measured data, the cost function can be written in the form, cf.~\cite{Cullimore:98}
\begin{eqnarray}\label{eq:cost}
&&S({\bf c}) = \sum_{n=0}^{N}\sum_{k=1}^{K}\left[ \eta_{T_w}\left({T}^k_{w,n}({\bf c})-\hat{T}^k_{w,n}\right)^2 +\right. \\\nonumber
&&~ \left. \eta_{T_f}\left({T}^k_{f,n}({\bf c})-\hat{T}^k_{f,n}\right)^2+\eta_{p} \left({p}^k_{n}({\bf c})-\hat{p}^k_{n}\right)^2\right],
\end{eqnarray}
where $\eta_i$ are weighting coefficients for different types of measurements, index $k$ runs through different locations of the sensors, and the index $n$ corresponds to discrete time instants $t_0,...,t_N$.

\subsection{Sensitivity analysis}
\label{ss:sensitivity}

Once the objective function of optimization is chosen we proceed with the analysis of input-output relations for the model to determine the most sensitive model parameters. At this step we evaluate how much each model parameter is contributing into the model uncertainty.  
\begin{figure}[h!]
	\centering
	\includegraphics[width=\linewidth,height=3.5cm]{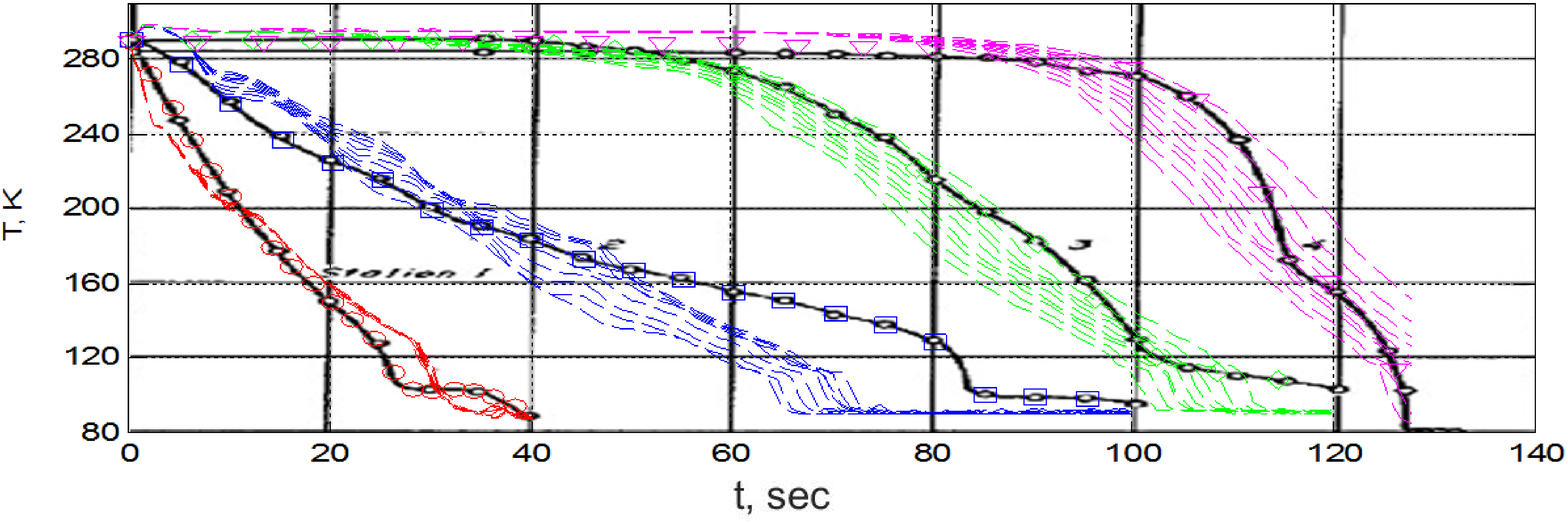}\\
	\vspace{-0.25cm}
	\caption{Results of the sensitivity analysis for $Gwsc$ experimental time-series data obtained at NIST. Data recorded at different locations are shown by black solid lines for fluid temperature at three locations. Colored dashed lines show model predictions at: (red) 0.6m from the entrance; (blue) 24 m; (green) 43 m; (pink) 60 m }\label{fig:sensitivity}
\end{figure}

We perform this test for each motel parameter. An example of the test outcome for overall scaling coefficient for mass transfer at the wall $Gwsc$ is shown in Fig.~\ref{fig:sensitivity}.  The sensitivity can be estimated as the relative change of the cost function normalized by the relative change of the parameter.  In this particular example 12 \% change in the parameter value results in 88 \% change in the cost function, i.e. the sensitivity is considered to be very high, except for the data obtained at station 1.

The results of the sensitivity test were used primarily to simplify the model by fixing parameters that have no effect on the output and to rank the most sensitive parameters and to learn their effect on the output of the model at various sensor locations. Typically it was found that only 20 parameters can be retained for subsequent model calibration. 

\subsection{Direct search}
\label{ss:search}

At the first step of the model calibration we used a simplified direct search to determine roughly the values of globally optimal model parameters.

\begin{figure}[h!]
	\centering
	\includegraphics[width=\linewidth,height=3.5cm]{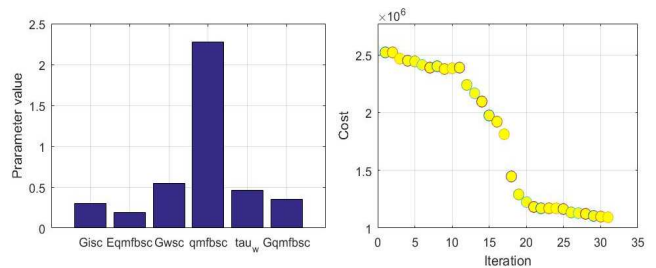}\\
	\vspace{-0.25cm}
	\caption{(a) Estimated values of the model parameters. (b) Convergence of
		the simplified direct search algorithm for simultaneous optimization of
		6 model parameters. }\label{fig:direct_search}
\end{figure}

Simplified direct search algorithm developed in this work has proven to be highly efficient at this stage. The algorithm is searching for a minimum of the cost function on a regular grid in multi-dimensional parameter space by scanning one  parameter at a time. The search is repeated several times with randomly changing order of the scanning directions. The convergence of the algorithm is illustrated in the Fig.~\ref{fig:direct_search}.

In this example the following 6 parameters were analyzed: scaling coefficients of the mass transfer at the interface (Gisc) and at the welted wall (Gwsc), characteristic time of the heat transfer to the wall (tauw), scaling coefficient for the film boiling heat transfer (qmfbsc), coefficients $c_2$ (Gmfbsc) and $c_3$ (Emfbsc) in eq. (\ref{eq:Iloeje_hfb}) for correction of the minimum heat flux.

The main advantage of this algorithm is that allows to determine quickly an approximate location of the global minimum in a given subspace of parameter space for poorly defined  initial guess. tie: example, the algorithm can scan within one hour uo to 30 parameters of the NIST model using 10 different scanning orders.

An approximation to the values of the model parameters found at this step can be further refined using one of the global optimization algorithms. 

\subsection{Global optimization}
\label{ss:optimization}

We note that casting the problem of fitting model predictions for two-phase flow in the standard form (\ref{eq:SSM}), (\ref{eq:cost}) allows one to use any available standard library for the solution of the optimization problem. In this work we performed global  optimization using a set of optimization algorithms available in MATLAB. We have verified the convergence of the model predictions towards experimental time series using pattern search, genetic algorithm, simulated annealing, and particle swarm algorithms. 
\begin{figure}[h!]
	\centering
	\includegraphics[width=7.cm]{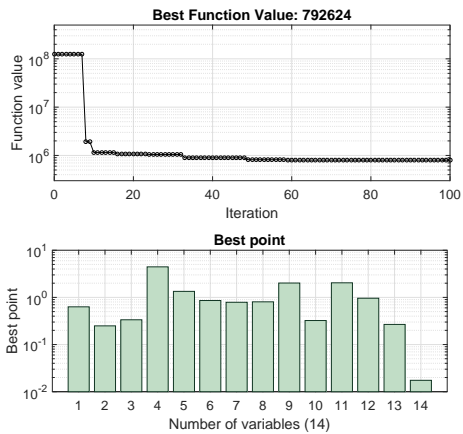}\\
	\vspace{-0.5cm}
	\caption{(a) Convergence of the simulated annealing algorithm for simultaneous optimization of 14 model parameters. (b) Best values of the model parameters. }\label{fig:SA-conv}
\end{figure}

The convergence of the model predictions using simulated annealing algorithm is illustrated in  Fig.~\ref{fig:SA-conv}. We note that convergence is achieved for simultaneous optimization of 14 parameters of the model. Besides 6 parameters listed above the following parameters were added to simultaneous optimization: scaling for the the Ditus-Boetler exponents in the heat transfer correlations on both sides of liquid vapor interface (hgi0esc and hli0esc) and at the dry wall (hg0esc), overall scaling for the heat transfer to the dry wall (hg0sc) and to the interface on the gas side (hgisc), scaling for the temperatures of the critical heat flux (Tchfsc) and minimum film boiling (Tmfbsc), and parameters of the transition boundary to the dispersed flow regime (xmin). 

Once the estimation of the optimal values of the model parameters are refined  we can formally complete inference procedure by estimating the variance of the model parameters. 

\subsection{Variance of the model parameters}
\label{ss:variance}

To estimate variance we repeat optimization using local search with multiple restarts in the vicinity of the quasi–optimal parameter value.
Essentially, at this stage we enhance original sensitivity analysis using simplex algorithm. 
\begin{figure}[h!]
	\centering
	\includegraphics[width=7.5cm]{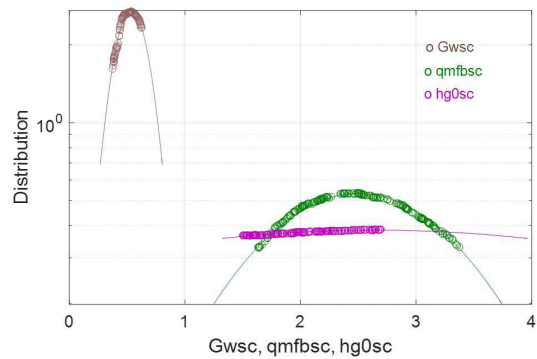}\\
	\caption{Example of estimation of the variances of the parameter values obtained using local search with multiple restarts.}\label{fig:dispersion}
\end{figure}

An example of estimation of the variance of parameter value is shown in the Fig~\ref{fig:dispersion}. In this example the distribution function for the parameter values obtained by direct calculations of the cost function for various values of the model parameter close to its optimal value are shown in figure by open symbols. 
The results of the direct numerical estimation of the distribution of the model parameters were fitted by Gaussian function 
\begin{equation}\label{eg:dispersion}
F = A(c_0)\cdot \exp\left(-\frac{1}{2}(c-c_0)\frac{\partial^2 S(c_0)}{\partial c^2}(c-c_0)\right)
\end{equation}
The results of the fitting are shown in the figure  by thin solid lines. We note that the fit by Gaussian function is quite satisfactory close to the maximum of the distribution. However, numerical simulations also reveal strong deviations from Gaussian fit for some values of the parameters.  Specifically, analysis shows the range of parameter values were simulations diverge.

These results also provide enhanced sensitivity analysis. For example, for parameter corresponding to the scale of the mass transfer coefficient at the wall ($Gwsc$) the dispersion $\sigma^2\approx0.025$ indicated the fact that the value of this parameter can be determined quite accurately using optimization procedure. 
On the other hand, the dispersion of the scaling coefficient of the heat transfer from the gas to the wall in the regime of forced convection ($hg0sc$) is very large $\sigma^2\approx12$, indicating that this parameter value can not be estimated accurately during optimization. 

The described optimization procedure is robust and sufficiently fast. Simultaneous optimization of 14 model parameters for the NIST (see next section) model with 30 control volumes, including sensitivity analysis, direct search, and global optimization can be computed in several hours on the laptop. 

Importantly, the proposed approach allows one to cast the fitting problem within a general inferential framework. Indeed, we begin with initial guess followed by rough estimation of  the distribution of the model parameters and then we use available experimental time-series data to update these distributions by estimating globally optimal values of the model parameters and their variance. This procedure can be systematically continued as soon as new experimental data become available. Furthermore, the approach can encompass comparison  between various alternative functional forms for two-phase flow correlations using time-series data available in multiple databases.

Using this approach we were able to demonstrate convergence of the model predictions towards experimental time-series obtained for chilldown of the cryogenic transfer lines under various experimental conditions ~\cite{Luchinsky2015c,Luchinsky2015a,Luchinsky2015b,Luchinsky2015d,LuchI:16d}. An example of such convergence is provided in the next section.

\section{Application to cryogenic transfer line} 
\label{s:transfer}

To validate this approach we used a set of experimental data obtained for chilldown in horizontal transfer line at National Bureau of Standards (currently NIST)~\cite{NBS:66} and chilldown large scale experimental transfer line at KSC~\cite{Robert2012}. Here we describe the result of the application of our approach to the analysis of chilldown in NIST experiment.

In the chilldown experiment~\cite{NBS:66} the vacuum jacketed line was 61 m long. The internal diameter of the copper pipe was 3/4 inches. Four measurement stations were located at the distance 6, 24, 42, and 60 m from the input valve. Three particular experimental data  sets were considered in this work: (i) subcooled liquid nitrogen and pressure in the storage tank was 4.2 atm; (ii) saturated liquid nitrogen flow driven by 3.4 atm pressure in the tank; and (iii) saturated liquid nitrogen flow driven by 2.5 atm pressure in the tank.

This set of experiments was selected for our analysis because it possesses a well-known difficulty for modeling, see e.g.~\cite{Cullimore:98}.

\subsection{Sub-cooled flow}
\label{ss:subcooled}

The results of modeling chilldown of cryogenic transfer line with sub-cooled liquid nitrogen flow under tank pressure 4.2 atm are shown in the next four figures.  The corresponding time-series data include fluid and wall temperature, the heat flux coefficient, and fluid pressure.

 \begin{figure}[h!]
 	\centering
 	\includegraphics[width=7.5cm]{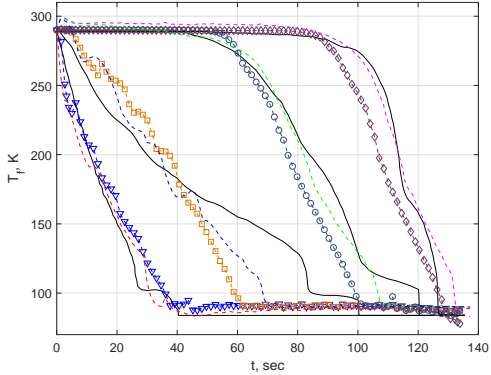}\\
 	\caption{Comparison of the model predictions (dashed colored lines) with the experimental time-series data (solid lines) for the fluid temperature measured at four locations along the pipe. Dashed colored lines and lines with colored open symbols correspond to the model predictions with two different sets of parameters. }\label{fig:Subcooled-Tf}
 \end{figure}

\begin{figure}[h!]
 	\centering
 	\includegraphics[width=7.5cm]{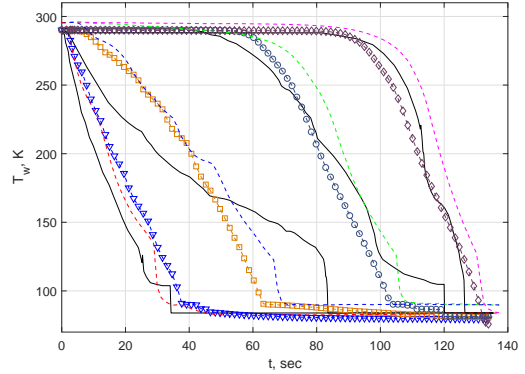}\\
 	\caption{Comparison of the model predictions (dashed lines) with the experimental time-series data (solid lines) for the wall temperature measured at four locations along the pipe. Color codding is the same as in previous figure.}\label{fig:Subcooled-Tw}
 \end{figure}

The results of comparison of the model predictions with the experimental data for the fluid temperature are shown in the Fig.~\ref{fig:Subcooled-Tf}. The corresponding comparison for the wall temperature is shown in the Fig.~\ref{fig:Subcooled-Tw}

Three different regions can be noticed in the figure. A fast cooling region in the beginning of the pipe. A region near the second station with long characteristic cooling time (order of 100 sec). And a region in the second half of the pipe that the remains hot for an extended period of time.

It can be seen from the figure that all three regions are reproduced by the model quite accurately both for the fluid and wall temperature. 

\begin{figure}[h!]
 	\centering
 	\includegraphics[width=7.5cm,height=4.5cm]{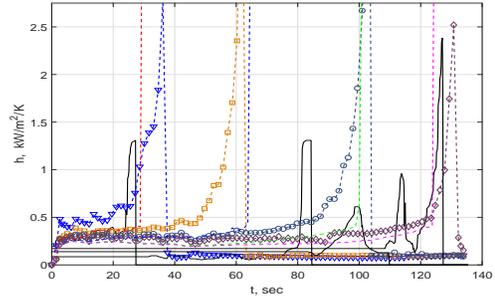}\\
 	\caption{Comparison of the model predictions (dashed lines with open symbols) with the experimental time-series data (solid black lines) for the heat transfer coefficient measured at four locations along the pipe.  Color codding is the same as in previous figures.  }\label{fig:Subcooled-htc}
 \end{figure}

\begin{figure}[h!]
	\centering
	\includegraphics[width=1.05\linewidth,height=5cm]{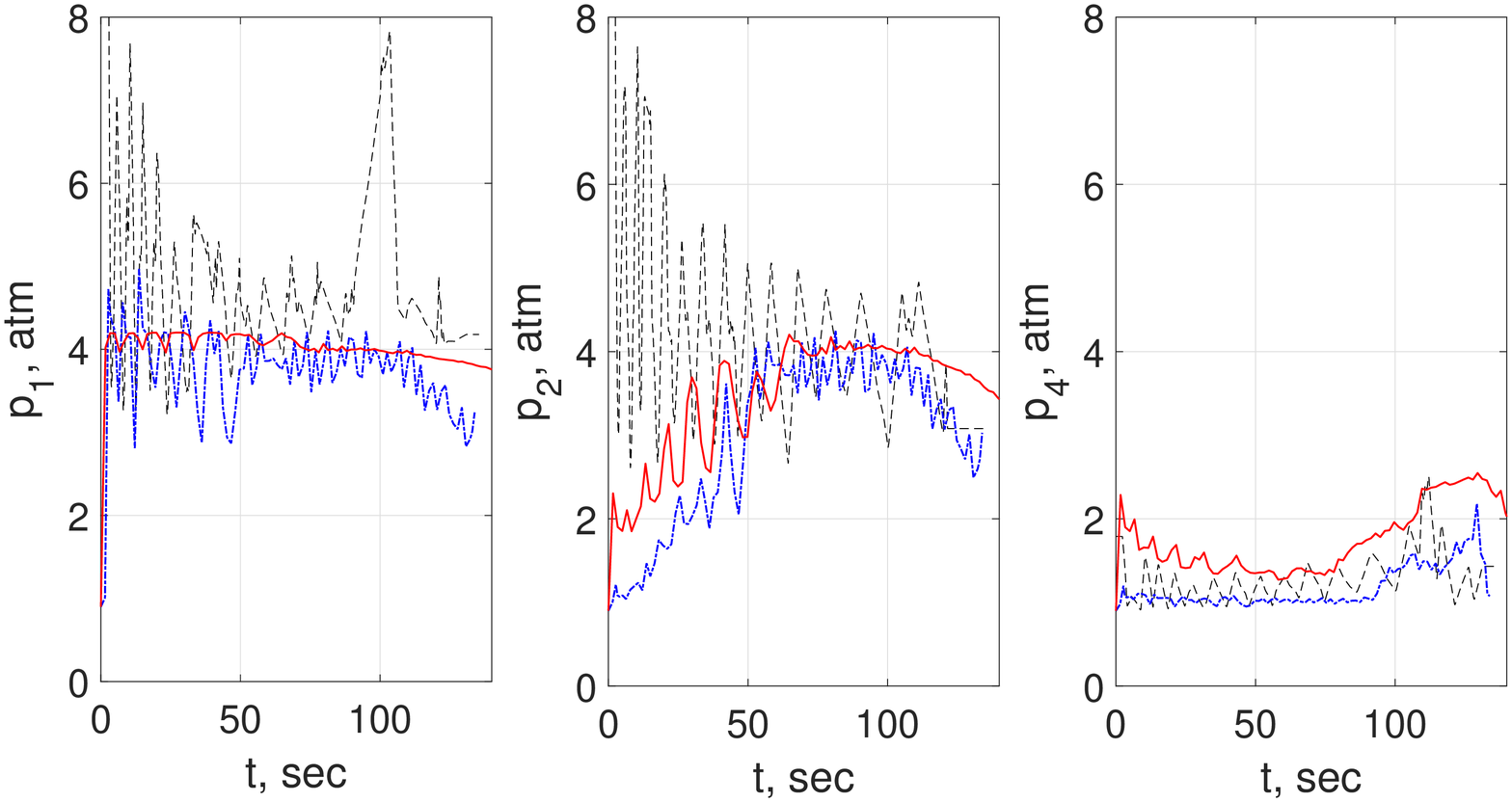}\vspace{-.5cm}
	\caption{Comparison of the model predictions obtained for two different sets of the model parameters  (blue dashed-dotted and red solid lines) with the experimental time-series data (dashed black lines) for the pressure measured at three locations along the pipe. Solid colored lines correspond to the model predictions with a different set of parameters.  }\label{fig:Subcooled-P}
\end{figure}

In general the solution of the optimization problem is not unique. Given different initial conditions the algorithm may converge to a slightly different values of parameters. Example of such convergence to two different sets of parameter values is illustrated in Figs.~\ref{fig:Subcooled-Tf} to \ref{fig:Subcooled-P} by different color codding. 

Both sets of parameters converged to the experimental time-series data within accepted tolerance and correspond to sub-optimal values of the cost function (\ref{eq:cost}). 

The non-uniqueness of the solution is a generic feature of the two-phase flow models that stems from the complex landscape of the cost function with multiple local minima. Regularization of the solution can be achieved e.g. by measurements of the additional flow variables or by testing the flow under different flow conditions. 

For example, the comparison of the model predictions with experimental time-series for the heat transfer coefficient and for the pressure are shown in Fig.~\ref{fig:Subcooled-htc} and Fig.~\ref{fig:Subcooled-P} respectively. 

It can be seen from the figures that experimentally estimated values of the total heat transfer coefficient to the wall are nearly constant at all locations and times except for a few narrow peaks. Therefore, the analysis of the heat transfer coefficient can provide in this case only semi-quantitative validation of the model predictions.

The comparison of the model predictions with experimental time-series data for the pressure shown in Fig.~\ref{fig:Subcooled-P} (note that the pressure time-series data are available only at three locations) are more informative. The model can capture semi-quantitatively the frequency and the mean values of the pressure  oscillations. 

However, large amplitude oscillations of pressure signal cannot be reproduce by the model. The most likely reason for this discrepancy is the dynamics of the input valve, which parameters are unknown. Therefore, during numerical experiments we usually limited contribution of the pressure signal to the cost function by setting values of $\eta_p$ to $\sim 0.1$ in eq.  (\ref{eq:cost}). 

 

\subsection{Saturated flow}\label{ss:saturated_flow}
	
As was mentioned above the convergence of the may be further improved by extending analysis to encompass time-series data obtained under different flow conditions. Following this idea we have included into our analysis the time-series data obtained in NIST experiment~\cite{Berenson:61} for saturated flows for two different driving pressures in the storage tank. Here we consider chilldown in the horizontal line observed for saturated nitrogen flow driven by the tank pressure 3.4 atm, see Fig.~\ref{fig:Subcooled-Tf}.
 \begin{figure}[h!]
 	\centering
 	\includegraphics[width=7.5cm]{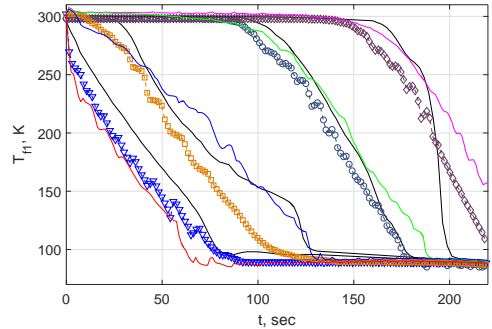}\\
 	\caption{Comparison of the model predictions (dashed colored lines) with the experimental time-series data (solid lines) for the fluid temperature measured at four locations along the pipe.  The nitrogen was under saturated conditions in the tank with pressure 3.4 atm.}\label{fig:Satur-34-atm}
 \end{figure}

It can be seen from the figure that the main effect of the reduced tank pressure (and corresponding reduction of nitrogen mass flow rate through the inlet valve) is an increase of the chilldown time by approximately 70 sec. Note, that the shape of the temperature signals remains essentially the same, cf. Fig.~\ref{fig:Subcooled-Tf}.

A good agreement between model predictions and experimental time-series data can be obtained using the same sets of the model parameters discussed above with small ( within 10\% ) adjustment of parameter $ tauw $. 
Similar results are obtained for saturated nitrogen flow under tank pressure 2.5 atm. 

	
We note, however, that the uncertainty in the inference of model parameters could not be resolved. We believe that the main reason for this is threefold: (i) the complexity of the temperature dynamics at the location of the 2-nd measurement station; (ii) the limited set of correlations adopted in this work for modeling cryogenic flow boiling during chilldown; and (iii) the limited information about system dynamics available in NIST time-series data. All these issues will be addressed in the future work in more details.

\nomenclature[]{$u$}{fluid velocity }%
\nomenclature[]{$p$}{pressure }%
\nomenclature[]{$T$}{temperature}%
\nomenclature[]{$E$}{total specific energy }%
\nomenclature[]{$H$}{total enthalpy}%
\nomenclature[]{$\dot q$}{heat flux}%
\nomenclature[]{$h$}{heat transfer coefficient}%
\nomenclature[]{$Re$}{Reynolds number }%
\nomenclature[]{$Gr$}{Grashoff number}%
\nomenclature[]{$Pr$}{Prandtl number}%
\nomenclature[]{$Pr$}{Prandtl number}%
\nomenclature[G]{$\alpha$}{gas void fraction}%
\nomenclature[G]{$\beta$}{liquid void fraction}%
\nomenclature[G]{$\Gamma$}{mass flow rate}%
\nomenclature[G]{$\rho$}{density}%
\nomenclature[G]{$\sigma $}{surface tension}%
\nomenclature[G]{$\tau$}{shear stress}%
\nomenclature[U]{$n$}{index for the time step}%

\section{Conclusion}
\label{s:conclusion}
	
To summarize, we developed fast and reliable solver for separated two-fluid cryogenic flow based on nearly-implicit algorithm and proposed a concise set of cryogenic two-phase flow boiling correlations capable of reproducing a wide range of experimental time-series data. 

The main emphasis in this work were placed on development of an efficient algorithm for simultaneous learning of a large number of parameters of cryogenic correlations that could ensure convergence of the model predictions towards experimental time-series data. 

Such an algorithm was proposed within inferential probabilistic framework. It involves the following steps: (i) sensitivity analysis of the model parameters, (ii) simplified direct search for approximate globally optimal values of these parameters, (iii) global stochastic optimization that refines the estimate for parameter values obtained at the previous step, and (iv) estimation of variance of the model parameters using local non-linear optimization.

The proposed approach was used to analyze chilldown in the horizontal transfer line with liquid nitrogen flow. It was shown that the algorithm can reliably converge towards experimental time-series data in the space of $\sim$20 model parameters both for sub-cooled and saturated flows. 

At the same time the analysis revealed the non-uniqueness of inferred set of model parameters. The latter results indicates that to obtain more accurate and reliable predictions the set of correlations will have to be extended and validated on a larger database of experimental data. These issues will be addressed in the future work.  

Another direction of future research will involve development an automation of the proposed approach using machine learning framework. 

It is important to note that the machine learning approach will most likely underly autonomous control and fault management of two-phase flows in the future space missions. Therefore, its development may accelerate and improve both learning required correlation parameters and reliable design of future exploration missions relying on two-phase flow management in space.

\section*{Acknowledgments}

This work was supported by the Advanced Exploration Systems and Game Changing Development programs
at NASA HQ.  



\section*{References}
 \bibliographystyle{apsrev4-1}
 \bibliography{TPF_BIB}



%
\end{document}